%% file: main.tex
\documentclass[conference]{IEEEtran}
\IEEEoverridecommandlockouts

\usepackage{cite}
\usepackage{amsmath,amssymb,amsfonts}
\usepackage{graphicx}
\usepackage{textcomp}
\usepackage{xcolor}
\usepackage{booktabs}
\usepackage{tabularx}
\usepackage{array}
\usepackage{colortbl}
\usepackage[edges]{forest}
\usepackage{url}
\usepackage{placeins}
\usepackage{microtype}

\graphicspath{{IEEE-conference-template-062824/figures/}}

\def\BibTeX{{\rm B\kern-.05em{\sc i\kern-.025em b}\kern-.08em
    T\kern-.1667em\lower.7ex\hbox{E}\kern-.125emX}}

\begin{document}

\input{IEEE-conference-template-062824/frontmatter/title-and-abstract}
\input{IEEE-conference-template-062824/sections/introduction}

\input{IEEE-conference-template-062824/sections/taxonomy}
\input{IEEE-conference-template-062824/sections/benchmarks}
\input{IEEE-conference-template-062824/sections/comparative-analysis}

\input{IEEE-conference-template-062824/sections/challenges}

\input{IEEE-conference-template-062824/sections/conclusion}

\input{IEEE-conference-template-062824/backmatter/references}

\end{document}

%% file: IEEE-conference-template-062824/frontmatter/title-and-abstract.tex
\title{Reasoning--Search--Augmented LLMs: An Operational Taxonomy, Benchmark Analysis and Controlled Evidence Synthesis}

\author{
\IEEEauthorblockN{Biswas Poudel\textsuperscript{1,*},
Nilson Chapagain\textsuperscript{1},
Amit Kumar\textsuperscript{2}, and
Xianshun Jiang\textsuperscript{1}}
\IEEEauthorblockA{\textsuperscript{1}Texas A\&M University, College Station, TX, USA\\
\{bpoudel,nchapagain,xianshun\}@tamu.edu}
\IEEEauthorblockA{\textsuperscript{2}Texas A\&M University--Corpus Christi, Corpus Christi, TX, USA\\
akumar3@islander.tamucc.edu}
\thanks{\textsuperscript{*}Corresponding author.}
}

\maketitle

\begin{abstract}
Large language models (LLMs) demonstrate impressive emergent reasoning capabilities, yet they remain constrained by fixed model parameters and finite context windows. Even when augmented with static retrieval, they can produce statements that are unsupported by, or even contradictory to, retrieved evidence. This leads to hallucinations, difficulty incorporating up-to-date information, and failures to correctly interpret or reconcile sources. Recent work addresses these limitations by tightly coupling step-by-step reasoning with interactive search, treating retrieval as an action within an iterative agentic loop rather than a one-shot preprocessing step. This study reviews frameworks that implement such reasoning-search loops by interleaving planning, querying, retrieval, integration, verification, and stopping. We analyze these systems through four architectural axes: retrieval coupling, controller design, knowledge source, and grounding strictness. We separately treat the learning signal as a cross-cutting dimension because prompting, supervision, preference optimization, and reinforcement learning can instantiate the same architecture while producing materially different search, stopping, and grounding behavior. We further distinguish benchmark difficulty from RSA-loop demand and synthesize controlled within-study ablations and quality--cost deltas, avoiding invalid pooling across heterogeneous model backbones, retrievers, corpora, and evaluation protocols. Our goal is to clarify the landscape of reasoning-search-augmented LLMs and outline a unified perspective on the design choices shaping future research.

\end{abstract}


%% file: IEEE-conference-template-062824/sections/introduction.tex
\section{Introduction}
\label{sec:intro}

Large language models (LLMs) have demonstrated strong emergent reasoning capabilities across mathematics, programming, and open-domain question answering \cite{wei2022chain,kojima2022zeroshot}. However, they remain constrained by (i) static parametric knowledge fixed at training time \cite{petroni2019lmkb,lewis2020rag} and (ii) finite context windows \cite{liu2024lostmiddle}. As a result, LLMs may hallucinate or omit critical facts on complex, long-horizon, or time-sensitive queries \cite{lin2022truthfulqa,liu2024lostmiddle}. Chain-of-thought and self-consistency prompting improve intermediate reasoning, yet still operate primarily over the model's parametric memory \cite{wei2022chain,wang2023selfconsistency,besta2025demystifying}.

Two lines of work have partially addressed these limitations. Early tool-use methods such as ReAct and Self-Ask interleave natural-language reasoning with calls to search APIs, but often rely on hand-crafted prompts and heuristics to decide when to search and what to retrieve \cite{yao2023react,press2023measuring}. In parallel, retrieval-augmented generation (RAG) grounds generation in retrieved documents and reduces hallucinations compared to ungrounded LLMs \cite{fan2024surveyragmeetingllms}. Nevertheless, standard RAG is typically a one-shot pipeline and is vulnerable to retrieval irrelevance, poor evidence use, and latency bottlenecks \cite{barnett2024seven}. These limitations have motivated systems that treat search as a first-class, iterative action within the reasoning process.

In many reasoning-intensive settings, retrieval is not only about finding relevant text: evidence may be structured, attributed, and distributed across heterogeneous sources such as documents, knowledge bases, databases, and tool outputs. This makes reasoning-search systems more than simple retrieve-then-generate pipelines, since query formulation, evidence integration, and verification must often operate over constrained interfaces and partially structured observations. Several systems reason over graph-structured state spaces or traverse knowledge-graph paths, making intermediate evidence more inspectable and multi-hop reasoning more explicit (e.g., Think-on-Graph, ReKG-MCTS, GraphReader) \cite{sun2024tog,song-etal-2025-rekg-ReKG-MCTS,li-etal-2024-graphreader}. Likewise, tool- and knowledge-base-mediated retrieval turns querying into an explicit action over external environments, where results come from typed or otherwise constrained interfaces such as APIs, databases, and knowledge graphs (e.g., MRKL, Toolformer) \cite{karpas2022mrkl,schick2023toolformer}. These developments reinforce a broader view of retrieval as controlled evidence acquisition within the reasoning process rather than a fixed preprocessing step.

Building on this motivation, a new generation of systems learns search behavior as part of the reasoning process. For instance, Search-R1 uses reinforcement learning (RL) to train LLMs to generate multiple queries during step-by-step reasoning and learn a stopping policy, improving over static RAG on open-domain and multi-hop QA \cite{searchr1}.

These developments motivate the present study of reasoning--search-augmented large language models (RSA-LLMs). We formalize an RSA-LLM as a system in which external evidence acquisition is embedded within the model's internal reasoning and control loop, rather than being treated as an offline or fixed preprocessing stage. A system is classified as an RSA-LLM if and only if it satisfies the following three necessary conditions:

\begin{enumerate}
    \item \textbf{State-dependent evidence acquisition.} At least one decision governing evidence acquisition is conditioned on the current internal reasoning state and/or accumulated evidence. Such decisions include whether or when to retrieve, how to formulate or reformulate a query, which external source or tool to invoke, and when to terminate retrieval.
    \item \textbf{Evidence-bearing external interaction.} The system must obtain task-relevant information through interaction with an external information environment, including but not limited to web search engines, document corpora, structured knowledge bases, databases, or API-based evidence retrieval services.
    \item \textbf{Closed-loop evidence integration.} Retrieved evidence is not merely appended to the context but is integrated into the system's evolving internal state, thereby influencing subsequent reasoning steps, verification processes, answer construction, and/or future evidence-acquisition decisions.
\end{enumerate}

From an operational perspective, RSA-LLMs instantiate a recurrent plan--query--retrieve--integrate--verify--stop computation cycle. Planning identifies missing or uncertain information required for task completion; querying operationalizes these information needs into executable retrieval requests; retrieval obtains external evidence; integration updates the model's working memory or latent reasoning state; verification evaluates evidential adequacy and claim support; and stopping determines whether the system should terminate with a response or continue further search. These components should be interpreted as functional abstractions rather than strictly modular architectural components, and may be implemented implicitly within a unified controller. Importantly, a system may complete a task after a single retrieval step; however, it must support state-conditioned control decisions that allow continuation, query revision, verification, or termination based on observed evidence. Recent adaptive information-seeking frameworks such as AISO and interleaved reasoning--action paradigms such as ReAct exemplify this closed-loop structure, while more recent approaches including Search-R1 and ReSearch further demonstrate learned multi-turn retrieval and stopping policies \cite{zhu2021adaptive,yao2023react,searchr1,chen2025researchlearningreasonsearch}.

This definition delineates RSA-LLMs from closely related but structurally distinct system classes. In particular, classical retrieval-augmented generation (RAG) and Fusion-in-Decoder (FiD) architectures incorporate external evidence but operate under a fixed or predetermined retrieval schedule, and therefore function as static conditioning mechanisms rather than adaptive reasoning--search systems; accordingly, they are treated as reference baselines rather than RSA-LLMs \cite{lewis2020rag,izacard2021fid}. Similarly, repeated or multi-step retrieval pipelines remain outside the RSA-LLM class when retrieval actions are pre-specified and not conditioned on intermediate reasoning outcomes or observed evidence. Generic tool-use agents are included only insofar as tool invocations retrieve or access external, task-relevant evidence and participate in the closed-loop reasoning process; purely computational or procedural tool calls do not satisfy the evidence-acquisition criterion. Finally, internal structured reasoning mechanisms such as tree- or graph-based search over an already provided context do not constitute reasoning--search augmentation unless they explicitly trigger additional external evidence acquisition. Consequently, RSA-LLM membership is determined strictly by the three conditions above, whereas design dimensions such as controller architecture, reasoning topology, knowledge source, grounding constraints, and learning paradigm characterize implementation variants within the RSA-LLM class. To make the scope of the survey auditable, we additionally report the literature-search and selection procedure, distinguish core RSA systems from adjacent baselines and evaluation studies, and apply the operational inclusion criteria consistently across the reviewed literature.

Our contributions are: (i) an operational RSA-LLM definition and boundary; (ii) a four-axis system taxonomy with edge-case coding rules; (iii) a cross-cutting analysis of learning signals; and (iv) a structured review and evaluation framework connecting architectures with benchmarks, budgets, grounding mechanisms, and controlled within-study evidence.

\begin{figure*}[t]
    \centering
    \fcolorbox{black!30}{white}{
        \includegraphics[width=0.74\textwidth]{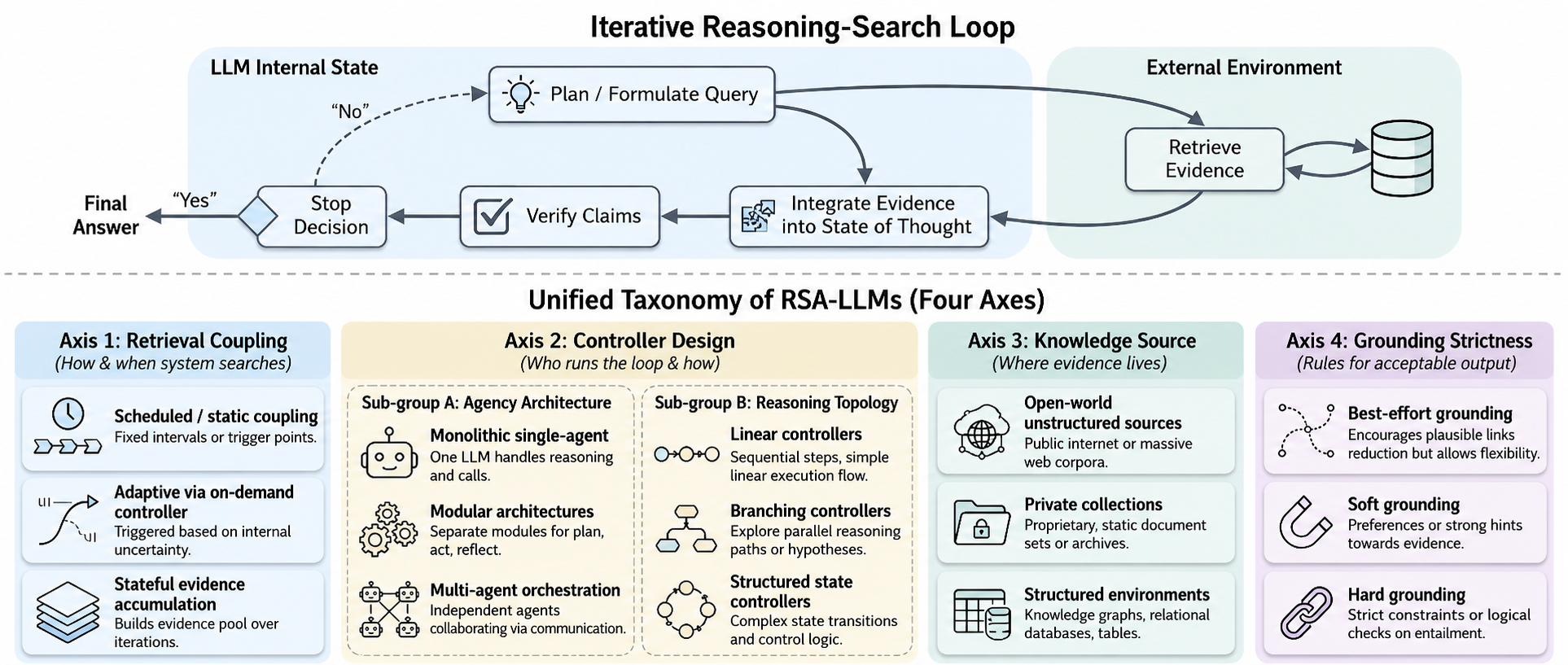}
    }
    \caption{Unified taxonomy of Reasoning–Search-Augmented LLMs. The top panel illustrates the iterative reasoning–search loop, while the bottom panel characterizes the design space along four orthogonal axes: retrieval coupling, controller design, knowledge source, and grounding strictness. Scheduled/Static retrieval is shown only as a reference baseline; learning signal is treated as a cross-cutting dimension.}
    \label{fig:unification_overview}
\end{figure*}

\subsection{Review Scope and Selection Protocol}

We followed a structured and auditable literature-selection protocol designed to make query construction and screening reproducible while providing representative coverage of reasoning-search-augmented LLMs. The search covered work published or publicly released between January 2020 and 31 May 2026. Candidate studies were identified through ACL Anthology, IEEE Xplore, ACM Digital Library, OpenReview, arXiv (cs.CL, cs.IR, cs.AI, and cs.LG), Semantic Scholar, and Google Scholar. For each source, we applied the Boolean template (“large language model” OR “LLM”) AND q, instantiating q separately with each of the following exact phrases: “reasoning with search,” “search-augmented reasoning,” “reasoning-search,” “adaptive retrieval,” “iterative retrieval,” “agentic RAG,” “deep research agent,” “multi-step retrieval,” “web-search agent,” and “reinforcement learning for search.” Database-specific syntax was adapted where necessary without changing the underlying search concepts. We supplemented these searches through backward and forward citation tracing from foundational and recent systems, including RAG, AISO, ReAct, Search-R1, and relevant RAG-reasoning surveys \cite{lewis2020rag,zhu2021adaptive,yao2023react,searchr1,li-etal-2025-survey}. During manuscript preparation, bibliographic metadata were updated only for studies already eligible by that date; no work first released after the cutoff was introduced.

Candidate studies were first screened by title and abstract and then assessed using the full text. The texts were evaluated against the operational conditions defined above. We excluded retrieval-only methods, fixed retrieve-then-generate pipelines, predetermined iterative retrieval, internal-only reasoning methods, and general tool-use systems whose tools did not acquire external evidence. Static RAG, retrieval components, and internal reasoning methods were retained only when they provided informative adjacent baselines. Feedback-space methods that search over internally generated artifacts using symbolic validator outputs are retained as adjacent control methods but excluded from the core RSA set when they do not acquire external task evidence; model-space search for planning-domain generation is one such case \cite{oswald2026modelspace}. Benchmark and evaluation studies were recorded separately when they measured retrieval, grounding, tool execution, or long-horizon search behavior relevant to RSA systems.

For each retained system, we extracted its publication status, learning signal, retrieval-coupling mechanism, controller architecture, reasoning topology, knowledge source, grounding mechanism, evaluation benchmarks, reported retrieval budget, and principal results. When multiple versions existed, we used the latest version available before the cutoff date and preferred peer-reviewed versions when available. Because of the rapid development of this area, relevant preprints were included but distinguished from peer-reviewed publications. The structured comparison set reported in Tables I and II comprises 13 core RSA systems, four adjacent systems, and 15 benchmark or evaluation studies. Additional studies were retained as component-level or contextual references but were not included in the system-level coding counts. The objective is transparent and reproducible coverage of representative design configurations rather than an exhaustive bibliometric census.

\subsection{Positioning Relative to Prior Surveys}

Existing surveys address several neighboring but distinct levels of analysis. General RAG surveys organize systems around retrieval, augmentation, generation, training, and application pipelines \cite{fan2024surveyragmeetingllms}. RAG-reasoning surveys primarily distinguish how reasoning improves retrieval-augmented generation, how retrieved knowledge supports reasoning, and how the two processes are synergistically combined \cite{li-etal-2025-survey}. Agentic-RAG surveys emphasize agent autonomy, orchestration patterns, reflection, planning, multi-agent collaboration, and practical workflow construction \cite{singh2025agenticrag}. Tool-learning surveys adopt the broader problem of teaching foundation models to select, invoke, and learn from external tools, regardless of whether those tools acquire evidence \cite{qin2024toollearning}.

Our scope is complementary but more operationally constrained. The unit of analysis is the closed-loop evidence-acquisition system: how the evolving reasoning state controls external search, how retrieved observations modify subsequent control, where evidence originates, and what evidential obligations govern the final output. The proposed taxonomy therefore contributes: (i) an explicit membership boundary separating RSA-LLMs from static RAG, internal-only reasoning, and non-evidential tool use; (ii) four jointly applicable system-level axes that separate acquisition control, controller organization, source constraints, and grounding obligations; (iii) reproducible coding rules for hybrid and boundary cases; and (iv) structured comparisons that connect architectural configurations with learning signals, benchmarks, retrieval budgets, grounding mechanisms, and reported outcomes. Unlike pipeline-centered or direction-of-influence taxonomies, this framing is designed to support system-level auditing and comparison of how evidence is acquired, controlled, and enforced during reasoning. Viewed as a data-systems problem, RSA-LLMs must search, integrate, and verify high-volume, heterogeneous, and rapidly changing evidence under token, latency, and API budgets. This connects retrieval control to scalable search, data integration, provenance, and cost-aware processing.

%% file: IEEE-conference-template-062824/sections/taxonomy.tex
\section{Taxonomy of Reasoning--Search--Augmented LLMs}
\label{sec:taxonomy}

Using the operational boundary established in Section~\ref{sec:intro}, we characterize qualifying systems along four architectural axes: retrieval coupling, controller design, knowledge source, and grounding strictness. The four axes are operational descriptors rather than statistically selected predictors of benchmark performance. They were chosen to separate independently variable system properties and to support reproducible coding; their empirical consequences are assessed only where matched comparisons are available. For comparative purposes, the retrieval-coupling axis also includes scheduled/static retrieval as an explicitly marked adjacent baseline that lies outside the RSA-LLM boundary. The remaining axes describe architectural design choices but do not independently fulfill RSA-LLM criteria. Learning signal is analyzed separately as the cross-cutting dimension introduced in Section II-E \cite{li-etal-2025-survey}.

Core or adjacent status is determined exclusively by the three membership conditions in Section~\ref{sec:intro}. Judge-, tree-, graph-, memory-, or tool-based mechanisms receive a core RSA label only when their evolving state changes subsequent external evidence acquisition or stopping. Hybrid systems may receive multiple architectural labels, while learning signal is coded separately from RSA membership.

% \vspace{-1.1em}
\subsection{Axis 1: Retrieval Coupling}
\label{sec:axis-coupling}

This axis describes \emph{how retrieval is invoked and controlled during inference}: when the system searches, what it queries, and when it stops searching.

\paragraph{\textbf{Adjacent reference point: scheduled/static retrieval}}

Scheduled/static retrieval occurs at predetermined points, with a fixed number of retrieval rounds or a predetermined top-$k$ evidence set. Classic RAG retrieves passages before generation, while FiD strengthens evidence integration by jointly attending over multiple retrieved passages without introducing a state-dependent retrieval controller \cite{lewis2020rag,izacard2021fid}. Even a multi-round pipeline remains scheduled when earlier observations do not alter later retrieval actions. Prior work on adaptive information seeking similarly distinguishes learned, evidence-conditioned retrieval policies from iterative strategies that repeatedly apply predefined retrieval operations \cite{zhu2021adaptive}. Scheduled systems are retained on the coupling axis to provide a common reference point, but they do not satisfy the RSA-LLM membership criterion.

\paragraph{\textbf{Adaptive coupling: the RSA boundary}}

Adaptive coupling begins at the RSA-LLM boundary. At least one retrieval trigger, query, source-selection, or stopping decision is conditioned on the evolving reasoning or evidence state. AISO, ReAct, Search-R1, and ReSearch instantiate this principle through learned or prompted policies that use previous reasoning and observations to determine subsequent actions \cite{zhu2021adaptive,yao2023react,searchr1,chen2025researchlearningreasonsearch}. ZeroSearch reduces training cost by simulating retrieval during training while transferring to real search at test time \cite{zerosearch} and StepSearch makes control more fine-grained through step-wise PPO with information-gain and redundancy-aware rewards \cite{wang2025stepsearch}. Other systems in this regime include R1-Searcher and Search-o1, which likewise vary when to retrieve, how to use evidence, and when to stop \cite{song2025r1searcher,li2025searcho1}. Recent search-and-refine systems introduce an explicit evidence-distillation state between retrieval rounds. AutoRefine and OThink-SRR1 filter or compress retrieved documents before continued reasoning, allowing the refined evidence state to influence subsequent querying, integration, and stopping decisions \cite{shi2025autorefine,liang2026othink}.

\paragraph{\textbf{Stateful adaptive coupling}}

Stateful coupling is a stronger form of adaptive coupling, rather than a mutually exclusive membership category, in which a persistent evidence, memory, or trajectory state conditions later retrieval and reasoning decisions. This changes coupling in a practical way: retrieval decisions become conditioned not only on the raw question and the latest snippets, but also on what has already been discovered and recorded \cite{wu2025resumunlockinglonghorizonsearch,memsearcher}. ReSum \cite{wu2025resumunlockinglonghorizonsearch} addresses long-horizon search under limited context by periodically summarizing the interaction history into a compact state that can support further exploration without overflowing the context window. MemSearcher \cite{memsearcher} similarly maintains and updates a compact memory after each search step, and trains the full loop end-to-end with RL. These designs improve stability in long search sessions, but they introduce new problems in evidence selection, compression, and drift: what gets written into the workspace can dominate later retrieval and reasoning, so pruning and update rules become first-class design choices \cite{wu2025resumunlockinglonghorizonsearch,memsearcher}. Related iterative RAG systems such as ReSP likewise make retrieval state explicit by recording and compressing the retrieval trajectory, reducing context overload while supporting continued multi-step evidence acquisition \cite{jiang2024resp}.

\subsection{Axis 2: Controller Design (Where control lives, and how inference is organized)}
\label{sec:axis-controller}

Controller design specifies \emph{who runs the loop} and \emph{how inference is structured}. We decompose this axis into two complementary sub-axes.

\paragraph{\textbf{Agency architecture (who runs the loop)}}
Single-agent designs place planning, retrieval, verification, and answering inside one loop; ReAct is a canonical prompt-based instance, while Search-R1 and Search-o1 are representative agentic search variants \cite{yao2023react,searchr1,li2025searcho1}. Modular architectures separate roles such as planner, retriever, and verifier, improving controllability but requiring coordination across modules and interfaces \cite{shen2023hugginggpt}. VerlTool exemplifies this setting by standardizing tool interaction and supporting modular training and evaluation across multiple tools \cite{jiang2025verltoolholisticagenticreinforcement}. BrowseMaster provides a recent open-web example of modular control in which a planner adaptively delegates retrieval subtasks to a programmatic executor and replans from the executor's distilled observations \cite{pang2025browsemaster}. Multi-agent orchestration distributes reasoning across specialized agents, often improving robustness at the cost of latency and aggregation complexity \cite{du2023multiagentdebate,liang2024mad}. PRISM is a retrieval-centric example that splits question analysis, evidence filtering, and evidence expansion across agents \cite{nahid2025prism}.

% \vspace{-1.1em}

\paragraph{\textbf{Reasoning topology (how inference is structured)}} Linear controllers follow one chain of reasoning, optionally inserting retrieval and verification steps; they are efficient but vulnerable to early errors that propagate through the remaining trace \cite{wei2022chain,yao2023tree}. Branching controllers explore multiple candidate paths to hedge against such errors, as in Tree-of-Thoughts and MCTS-style search \cite{yao2023tree,browne2012mcts}. Structured-state controllers maintain an explicit state representation such as a graph or belief state, improving traceability at the cost of additional state construction and pruning \cite{besta2023graphofthoughts}. Think-on-Graph, Graph-of-Thoughts-style methods, and GraphReader illustrate how graph-structured intermediate states can support controlled multi-hop reasoning and provenance tracking \cite{sun2024tog,jin-etal-2024-graph,li-etal-2024-graphreader}. CoT, ToT, MCTS, and graph-based reasoning are included here as topology archetypes; they constitute RSA controllers only when their evolving states trigger external evidence acquisition.

\subsection{Axis 3: Knowledge Source}
\label{sec:axis-knowledge}

This axis specifies where external evidence resides and the constraints that the source imposes on acquisition, integration, and verification. Knowledge source alone does not determine RSA eligibility: a system qualifies only when observations obtained from the source influence subsequent reasoning or retrieval actions.

\noindent\emph{Open-world unstructured sources:}
The open web and large public corpora provide broad and potentially current coverage, but expose the controller to source volatility, redundancy, conflicting evidence, and variable authority. WebGPT exemplifies this regime because its browser actions are conditioned on the evolving interaction history \cite{nakano2021webgpt}. Evaluation in this setting must therefore report the search service, access date, and retrieval budget, since changes in indexing or page availability can materially affect results.

\noindent\emph{Bounded or private collections:}
Enterprise repositories, domain-specific collections, and fixed corpus snapshots permit stable indexing, access control, and more reproducible evaluation, but their coverage and freshness are bounded by corpus maintenance. Search-R1 illustrates how state-dependent queries can be executed against a fixed local retrieval stack \cite{searchr1}. Such environments enable controlled comparisons, although performance may remain sensitive to corpus construction, retriever configuration, and document version.

\noindent\emph{Structured environments:}
Databases, knowledge graphs, and evidence-bearing APIs expose typed operations rather than unconstrained text search. Controllers must select an appropriate interface, generate valid calls, interpret structured responses, and recover from execution failures. KnowTrace illustrates this regime by converting retrieved information into an explicit knowledge trace that supports subsequent retrieval and reasoning \cite{li2025knowtrace}. Purely computational tool calls remain outside the RSA boundary unless their outputs constitute external evidence used by later control decisions.

\subsection{Axis 4: Grounding Strictness}
\label{sec:axis-grounding}

Grounding strictness characterizes how strongly a system constrains its output to retrieved evidence. Operationally, evidence may be merely available to the generator, encouraged through learning or revision signals, or required through an explicit support contract.

\noindent\emph{Best-effort grounding:}
Retrieved evidence informs generation, but individual claims need not be explicitly supported. Static RAG and FiD provide adjacent reference points for this regime: retrieved passages can improve answer quality, yet the generator may ignore evidence or introduce unsupported details \cite{lewis2020rag,izacard2021fid}. Best-effort grounding therefore provides neither claim-level attribution nor a prescribed response when sufficient evidence is unavailable.

\noindent\emph{Soft grounding:}
Unsupported outputs are discouraged through rewards, reflection, reranking, critique, or revision, but remain possible. Search-R1 and StepSearch optimize outcome- or retrieval-aware signals without imposing claim-level support constraints \cite{searchr1,wang2025stepsearch}. Self-RAG uses reflection tokens to regulate retrieval and critique, whereas RARR retrieves evidence to revise potentially unsupported statements \cite{asai2024selfrag,gao-etal-2023-rarr}. These mechanisms improve evidence use while preserving generation flexibility.

\noindent\emph{Hard grounding:}
A hard-grounded system requires claims to be supported, cited, revised, or withheld when adequate evidence cannot be established. GopherCite approximates this regime through attributable quotations and abstention under insufficient support \cite{menick2022gophercite}. ALCE operationalizes complementary citation-coverage and citation-correctness criteria for evaluating attributable generation \cite{gao-etal-2023-alce}. Hard grounding improves auditability, but may reduce answer coverage when retrieval fails or automated support judgments are unreliable.

\subsection{Cross-Cutting Dimension: Learning Signal}
\label{sec:learning-signal}

The four primary axes describe system architecture and inference-time behavior, whereas the learning signal describes how that behavior is elicited or optimized. These dimensions are architecturally separable: similar reasoning--search loops may be implemented through prompting, supervised trajectories, preference optimization, or reinforcement learning. Nevertheless, the learning signal is behaviorally consequential because it affects query formulation, source selection, evidence integration, stopping, and grounding. We therefore treat it as a cross-cutting analytic dimension rather than a fifth architectural axis.

We assign non-exclusive learning-signal tags to each reviewed system. Prompted or in-context control (P) elicits search behavior through instructions and demonstrations without parameter updates, as exemplified by ReAct \cite{yao2023react}. Self-supervised tool induction (SS) derives training targets from model-generated or automatically filtered interactions, as in Toolformer \cite{schick2023toolformer}. Supervised fine-tuning or imitation (SFT) learns from curated reasoning, search, or tool-use trajectories. Preference-based optimization (Pref) uses human or model preferences to favor helpful, attributable, or evidence-consistent behavior, as illustrated by WebGPT \cite{nakano2021webgpt}. Outcome-level reinforcement learning (RL-O) optimizes final-task reward and can induce multi-turn search without supervised reasoning traces, as in Search-R1 and ReSearch \cite{searchr1,chen2025researchlearningreasonsearch}. Process- or retrieval-aware reinforcement learning (RL-P) supplies intermediate signals such as information gain, evidence utility, redundancy penalties, or retrieval cost, as in StepSearch and R-Search \cite{wang2025stepsearch,rsearch}.

These signals introduce different control trade-offs. Prompting is inexpensive and portable but sensitive to model capability and prompt formulation. SFT and imitation stabilize action syntax and search trajectories but may reproduce demonstration biases and generalize poorly beyond observed workflows. Outcome-level RL permits strategy discovery but creates sparse credit-assignment problems because final correctness does not identify which retrieval actions were useful. Process-level rewards provide denser supervision, but their proxy metrics may reward excessive search, superficial evidence overlap, or other behavior that does not improve final grounding. Preference optimization can strengthen attribution and evidence adherence, although its behavior remains dependent on the coverage and reliability of the preference model.

Learning signal must therefore be reported when comparing RSA architectures. A performance difference attributed to retrieval coupling or controller design may instead arise from supervision quality, reward granularity, training data, or policy optimization. Accordingly, Table~\ref{tab:system-comparison} records learning signal alongside the four architectural axes, thereby preventing differences in supervision or optimization from being conflated with differences in system architecture.

%% file: IEEE-conference-template-062824/sections/benchmarks.tex
\begin{table*}[!t]
\caption{Representative benchmarks and their demand for an adaptive reasoning--search loop. L, M, and H denote low, moderate, and high RSA-loop demand; Var. denotes item-dependent demand.}
\label{tab:benchmarks}
\centering
\scriptsize
\setlength{\tabcolsep}{2.7pt}
\renewcommand{\arraystretch}{0.98}
\begin{tabular}{p{1.55cm} p{3.05cm} p{4.25cm} p{0.95cm} p{5.90cm}}
\hline
\textbf{Setting} & \textbf{Benchmark} & \textbf{Primary capability} &
\textbf{Demand} & \textbf{Main limitation} \\
\hline
Fact lookup &
NQ; TriviaQA \cite{kwiatkowski-etal-2019-natural-questions,JoshiTriviaQA2017} &
Open-domain retrieval and short-answer generation &
L &
Often solvable from one passage; static corpora and contamination limit
evaluation of search control. \\

Multi-hop QA &
HotpotQA; MuSiQue;
2WikiMultiHopQA
\cite{yang-etal-2018-hotpotqa,trivedi-etal-2022-musique,ho-etal-2020-2WikiMultiHopQA} &
Cross-document evidence composition and supporting-fact reasoning &
M &
Fixed or provided contexts permit pre-retrieval or long-context inference;
EM/F1 obscures retrieval control. \\

End-to-end RAG &
FRAMES \cite{krishna2025frames} &
Factuality, retrieval, and multi-constraint reasoning &
M &
Small, Wikipedia-centered set that does not isolate controller behavior. \\

Verification &
FEVER \cite{thorne-etal-2018-fever} &
Evidence retrieval, claim support, and veracity classification &
L &
Static Wikipedia and synthetic claims allow fixed retrieval without
adaptive search. \\

Noisy search &
SEALQA \cite{pham2026sealqa} &
Ambiguous, conflicting, time-sensitive, and distractor-heavy evidence &
H &
Small, version-sensitive set; LongSeal tests selection from supplied
documents rather than active acquisition. \\

State-gated retrieval &
SGR-Bench \cite{li2026sgrbench} &
Configuring filters, hierarchies, and scopes to expose answer-bearing evidence &
H &
Only 100 tasks; preprint status and changing interfaces limit stability. \\

Multimodal web &
MERRIN \cite{wang2026merrin} &
Multimodal evidence retrieval and reasoning under noisy or conflicting sources &
H &
High tool and media costs plus changing web content limit reproducibility. \\

General assistants &
GAIA \cite{mialon2024gaia} &
Planning, browsing, multimodal reasoning, and tool use &
Var. &
Mixes search with non-evidential tools; subset choice affects comparisons. \\

Deep browsing &
BrowseComp; BrowseComp-ZH
\cite{wei2025browsecompsimplechallengingbenchmark,zhou2025browsecompzh} &
Persistent search over entangled English and Chinese web clues &
H &
Accuracy omits evidence and citation quality; web drift and leakage remain
concerns. \\

Web traversal &
WebWalkerQA \cite{wu2025webwalker} &
Hierarchical website navigation, evidence accumulation, and stopping &
H &
Changing sites reduce reproducibility; scores conflate navigation,
retrieval, and generation. \\

Research synthesis &
DeepResearch Bench II \cite{li2026deepresearchbench2} &
Long-form evidence recall, analysis, and report presentation &
H &
Only 132 costly tasks; rubric scores conflate search, analysis, and writing. \\
\hline
\end{tabular}
\end{table*}
\section{Benchmarks and Evaluation}
\label{sec:evaluation}

Evaluation of RSA-LLMs requires separating benchmark difficulty from RSA-loop necessity. A dataset may require substantial factual knowledge or multi-step reasoning without testing state-dependent evidence acquisition: standard RAG, pre-retrieved documents, or a sufficiently long context may still be adequate. We therefore characterize benchmarks by the extent to which successful completion depends on adaptive querying, navigation, evidence integration, and stopping, in addition to summarizing the capability each benchmark measures.

\subsection{Benchmark Coverage and RSA-Loop Demand}
\label{sec:benchmark-coverage}

Table~\ref{tab:benchmarks} groups representative benchmarks into three levels of RSA-loop demand. Low-demand benchmarks primarily evaluate retrieval, answer generation, or verification and can generally be addressed using static retrieval. Moderate-demand benchmarks require distributed evidence or multi-hop reasoning, but their fixed corpora or provided contexts may permit predetermined retrieval or long-context inference. High-demand benchmarks make evidence discovery itself part of the task: subsequent queries, navigation actions, or source choices must typically be revised in response to previously observed evidence. \emph{Variable} denotes heterogeneous suites in which RSA-loop necessity depends on the selected subset.

This distinction prevents benchmark performance from being interpreted as direct evidence for the value of adaptive coupling. Natural Questions, TriviaQA, and FEVER primarily test retrieval or verification quality rather than search control. HotpotQA, 2WikiMultiHopQA, MuSiQue, and FRAMES require evidence composition, but do not necessarily force state-dependent acquisition when relevant documents can be retrieved in advance. By contrast, SEALQA, SGR-Bench, MERRIN, BrowseComp, and WebWalkerQA expose systems to noisy results, hidden retrieval states, multimodal sources, or evolving navigation paths for which later actions depend on earlier observations. Long-form suites such as DeepResearch Bench II additionally test whether retrieved evidence can support coherent, attributable synthesis.

\subsection{Metrics, Budgets, and Reproducibility}
\label{sec:metrics}

Final-answer correctness is necessary but insufficient because it does not identify whether failure arose from evidence acquisition, integration, grounding, or control. We therefore organize evaluation metrics into five complementary dimensions: (i) answer quality, measured by exact match, token-level F1, task-specific accuracy, or rubric-based judgments; (ii) retrieval quality, measured by evidence recall, precision, Recall@$k$, or supporting-fact F1; (iii) grounding quality, measured by citation precision and recall, attribution coverage, and entailment between claims and cited evidence; (iv) control and efficiency, measured by search calls, pages visited, retrieved and generated tokens, latency, monetary cost, redundancy, and stopping behavior; and (v) robustness, measured under distractor evidence, conflicting sources, tool failures, source drift, and repeated stochastic runs.

For reproducibility, each reported result should specify the base model and learning signal; corpus snapshot or search API and access date; retriever, reranker, and retrieval depth; maximum search calls or pages; context and token budgets; available tools; prompting protocol; evaluator or judge model and version; number of trials; and variance or confidence intervals where available. Architectural comparisons should use matched search and token budgets or report quality--cost Pareto curves. Results obtained using fixed corpora and live-web environments should not be pooled, and EM, F1, citation scores, and judge-based scores should not be treated as interchangeable effect sizes. Generic function-calling benchmarks may be retained as adjacent diagnostics of interface reliability, but they validate RSA behavior only when the invoked tools return task-relevant external evidence \cite{li-etal-2023-apibank,patil2025bfcl}.

For the quantitative synthesis in Section IV-C, we retained only contrasts reported within the same primary study. A comparison was treated as a component ablation when the study held the model backbone, retrieval environment, evaluation set, and principal inference budget fixed while removing or modifying a specified controller, state, or learning component. Comparisons that matched the backbone and evaluation environment but jointly changed multiple components were recorded separately as multi-factor contrasts. For EM, F1, and accuracy, effects are expressed as absolute percentage-point differences; relative changes are retained only when the primary study did not report the corresponding baseline score. When a study evaluated the same metric over several datasets, we computed an unweighted macro-average over exactly those datasets. Efficiency effects are reported as relative reductions in retrieval calls, processed tokens, or latency. Because most studies omit repeated-run variance or confidence intervals, we do not pool effects across papers, infer statistical significance, or treat heterogeneous metrics as interchangeable effect sizes.

%% file: IEEE-conference-template-062824/sections/comparative-analysis.tex
\begin{table*}[!t]
\caption{Representative systems across five design dimensions. Adjacent reference systems are excluded from cross-system counts.}
\label{tab:system-comparison}
\centering
\fontsize{7.5pt}{8pt}\selectfont
\setlength{\tabcolsep}{4.0pt}
\renewcommand{\arraystretch}{1.00}
\begin{tabularx}{\textwidth}{
@{}>{\raggedright\arraybackslash}p{2.05cm}
c
>{\raggedright\arraybackslash}p{2.55cm}
>{\raggedright\arraybackslash}X
c c@{}}
\toprule
\textbf{Method} &
\textbf{Coupling} &
\textbf{Controller} &
\textbf{Knowledge source / retriever} &
\textbf{Ground.} &
\textbf{Learn.} \\
\midrule

\rowcolor{black!7}
\multicolumn{6}{l}{\textbf{Panel A: Adjacent reference systems}} \\

RAG / FiD
\cite{lewis2020rag,izacard2021fid} &
Sch. &
Single / linear or fusion &
Wikipedia; DPR &
BE & SFT \\

Feedback-space planning
\cite{oswald2026modelspace} &
Ad. &
Search over candidate models &
VAL and landmark feedback &
N/A & P \\

GopherCite \cite{menick2022gophercite} &
Sch. &
Single / rerank & 
Open web; Google Search top-K &
Hard & SFT+Pref \\

\midrule
\rowcolor{black!7}
\multicolumn{6}{l}{\textbf{Panel B: Core RSA-LLMs}} \\

ReAct
\cite{yao2023react} &
Ad. &
Single / linear &
Wikipedia API &
BE & P \\

IRCoT
\cite{trivedi2023ircot} &
Ad. &
Single / linear &
Task-specific Wikipedia corpora; BM25 &
BE & P \\

Self-RAG
\cite{asai2024selfrag} &
Ad. &
Single / linear &
Contriever and web search; top-5 &
Soft & SFT \\

WebGPT
\cite{nakano2021webgpt} &
Ad. &
Single / linear &
Open-web browser &
Soft & SFT+Pref \\

ReKG-MCTS
\cite{song-etal-2025-rekg-ReKG-MCTS} &
St. &
Single / tree &
Freebase knowledge graph &
Soft & P \\

Search-R1
\cite{searchr1} &
Ad. &
Single / linear &
Wikipedia-2018; E5 &
Soft & RL-O \\

ReSearch
\cite{chen2025researchlearningreasonsearch} &
Ad. &
Single / linear &
Wikipedia-2018; E5 &
Soft & RL-O \\

StepSearch
\cite{wang2025stepsearch} &
Ad. &
Single / linear &
Wikipedia-2018; E5 &
Soft & RL-P \\

AutoRefine
\cite{shi2025autorefine} &
St. &
Single / linear-refine &
Wikipedia-2018; E5 &
Soft & RL-P \\

S2G-RAG
\cite{li2026s2grag} &
St. &
Modular / linear &
Dataset Wikipedia; BM25 or E5 &
Soft & SFT \\

OThink-SRR1
\cite{liang2026othink} &
St. &
Single / linear-refine &
Wikipedia-2023; Elasticsearch &
Soft & RL-P \\

BrowseMaster
\cite{pang2025browsemaster} &
St. &
Modular / hierarchical &
Open web; programmatic tools &
BE & P \\

PiCA
\cite{liu2026pica} &
Ad. &
Single / linear &
Wikipedia-2018; E5 &
Soft & RL-P \\

\bottomrule
\end{tabularx}

\vspace{1mm}
\begin{minipage}{0.99\textwidth}
\scriptsize
\textit{Notation:}
Sch.\ = scheduled/static;
Ad.\ = adaptive;
St.\ = stateful adaptive;
BE = best-effort grounding;
P = prompted or training-free control;
Pref = preference optimization;
RL-O = outcome-level reinforcement learning;
RL-P = process- or retrieval-aware reinforcement learning;
\end{minipage}
\end{table*}
\section{Comparative Performance and Trends}

\subsection{System-Level Comparison}
\label{sec:system-comparison}

Table~\ref{tab:system-comparison} maps representative systems onto the RSA
boundary, the four architectural axes, and the cross-cutting learning signal.
Adjacent references are shown separately and excluded from the descriptive
counts below. The set is representative rather than exhaustive and supports
taxonomy-level comparison, not performance ranking.

The mapping confirms that retrieval coupling and learning signal are
analytically distinct: adaptive or stateful acquisition appears under prompting,
supervised control, preference feedback, and both outcome- and process-aware
reinforcement learning \cite{yao2023react,li2026s2grag,nakano2021webgpt,
searchr1,wang2025stepsearch}. Controller comparisons must therefore account for
differences in supervision, reward design, and optimization. Feedback-space
planning is retained only as an adjacent boundary case because it adaptively
optimizes a symbolic artifact from validator feedback rather than acquiring
external task evidence \cite{oswald2026modelspace}.

\subsection{Cross-System Trends}

Among the 13 core systems, ten use single-controller linear or linear-refinement topologies, ten operate over fixed corpora or knowledge graphs, five maintain an explicit stateful retrieval or evidence representation, and six use outcome- or process-aware reinforcement learning. All 13 employ best-effort or soft grounding; GopherCite provides the hard-grounded adjacent reference. These descriptive counts should not be interpreted as bibliometric prevalence estimates.

Trends supported across several systems include the separation of adaptive retrieval from reinforcement learning, the movement from transient retrieval observations toward persistent evidence states, and increasing use of process-aware rewards for search control. By contrast, conclusions about branching controllers or planner--executor architectures remain setting-specific: ReKG-MCTS demonstrates their utility over knowledge-graph paths, while BrowseMaster does so in long-horizon open-web search \cite{song-etal-2025-rekg-ReKG-MCTS,pang2025browsemaster}. Such observations motivate controlled comparisons but do not, by themselves, establish that one controller family is generally superior.

\subsection{Controlled Quantitative Synthesis}

Direct pooling of published RSA-LLM scores would confound model scale, training data, retriever quality, corpus snapshot, benchmark subset, and search budget. Table~\ref{tab:controlled-evidence} therefore reports only controlled component ablations and explicitly identified matched comparisons extracted from the original studies. This design converts the comparative discussion from a cross-paper leaderboard into a set of localized effect estimates: each delta answers whether a particular component mattered under the study's own fixed evaluation conditions.

\begin{table*}[!t]
\centering
\caption{Controlled quantitative evidence. Quality entries give median absolute change [min, max], positive-effect counts, and $k_Q$ benchmark--metric observations. S, C, M, and A denote single-component, conditional, multifactor, and aggregate-only evidence; studies are not pooled.}
\label{tab:controlled-evidence}

\fontsize{7.5pt}{8pt}\selectfont
\setlength{\tabcolsep}{3.2pt}
\renewcommand{\arraystretch}{1.02}

\begin{tabularx}{\textwidth}{
@{}p{1.65cm}p{3.0cm}c c X p{2.25cm}p{1.0cm}@{}}
\toprule
Study &
Controlled contrast &
Type &
$k_Q$ &
Reported quality change &
Efficiency change &
Source \\
\midrule

\rowcolor{black!7}
\multicolumn{7}{l}{\textbf{Panel A: Single-component and module-level evidence}} \\

Search-R1 \cite{searchr1}
& Retrieved-token masking
& S & 14
& 7B EM: $+8.8$ [$+6.1,+12.8$], 7/7 $\uparrow$;
  3B EM: $+4.3$ [$-1.6,+10.3$], 5/7 $\uparrow$
& Not reported
& T6 \\

StepSearch \cite{wang2025stepsearch}
& Step-wise process reward
& S & 8
& EM: $+2.45$ [$+1.9,+8.3$], 4/4 $\uparrow$;
  F1: $+2.95$ [$+1.7,+8.8$], 4/4 $\uparrow$
& Not reported
& T4 \\

AutoRefine \cite{shi2025autorefine}
& Retrieval-specific reward
& S & 14
& Base EM: $+3.7$ [$0.0,+4.4$], 6/7 $\uparrow$;
  Instruct EM: $+4.5$ [$+1.0,+6.4$], 7/7 $\uparrow$
& Not reported
& T2 \\

AutoRefine \cite{shi2025autorefine}
& Refinement step under no retrieval reward
& S & 14
& Base EM: $+7.4$ [$-0.2,+20.0$], 6/7 $\uparrow$;
  Instruct EM: $+2.3$ [$+0.7,+6.2$], 7/7 $\uparrow$
& Not reported
& T2 \\

S2G-RAG \cite{li2026s2grag}
& Fine-tuned versus untrained judge
& S & 2
& HotpotQA: $+4.1$ EM; $+5.7$ F1
& Not reported
& T2 \\

S2G-RAG \cite{li2026s2grag}
& S2G-Judge versus no judge
& S & 2
& HotpotQA: $+15.8$ EM; $+18.9$ F1
& Not reported
& T2 \\

S2G-RAG \cite{li2026s2grag}
& Evidence extractor versus raw accumulation
& S & 2
& HotpotQA: $+3.8$ EM; $+4.0$ F1
& Latency $\downarrow 17.7\%$
& T2--3 \\

S2G-RAG \cite{li2026s2grag}
& Structured versus free-text gap query
& S & 3
& HotpotQA: $+1.8$ EM; $+2.0$ F1;
  $+2.2$ correct retrieval
& Not reported
& T6 \\

\midrule
\rowcolor{black!7}
\multicolumn{7}{l}{\textbf{Panel B: Conditional and multifactor architectural evidence}} \\

Browse\allowbreak Master \cite{pang2025browsemaster}
& Planner contribution
& C & 2
& BrowseComp: $+1.5$ without primitives;
  $+15.0$ with primitives
& Not reported
& T2 \\

Browse\allowbreak Master \cite{pang2025browsemaster}
& Programmatic-primitives contribution
& C & 2
& BrowseComp: $+5.5$ without planner;
  $+19.0$ with planner
& Not reported
& T2 \\

Browse\allowbreak Master \cite{pang2025browsemaster}
& Full planner--primitive system
& M & 1
& BrowseComp: $+20.5$ accuracy
& Not reported
& T2 \\

OThink-SRR1 \cite{liang2026othink}
& Search--refine--reason versus ReSearch
& M & 16
& 7B: EM $+1.26$ [$-0.52,+1.66$], 3/4 $\uparrow$;
  F1 $+0.36$ [$-0.80,+1.19$], 3/4 $\uparrow$.
  3B: EM $+3.84$ [$+2.53,+8.45$], 4/4 $\uparrow$;
  F1 $+5.88$ [$+4.30,+11.32$], 4/4 $\uparrow$
& 7B calls $\downarrow 31.2\%$ [28.0,35.5];
  tokens $\downarrow 44.1\%$ [36.0,52.6]
& T1/F4 \\

\midrule
\rowcolor{black!7}
\multicolumn{7}{l}{\textbf{Panel C: Aggregate-only evidence}} \\

PiCA \cite{liu2026pica}
& Pivot-based versus outcome-only credit
& A & --
& Across four model families:
  relative EM $+6.6\%$ to $+34.0\%$;
  relative F1 $+4.1\%$ to $+29.1\%$
& Not reported
& T3 \\

\bottomrule
\end{tabularx}

\vspace{1mm}
\begin{minipage}{\textwidth}
\scriptsize
\textit{Notes:}
Brackets denote the range across benchmarks, not confidence intervals.
For BrowseMaster, component effects are conditional on whether the
complementary component is present and therefore should not be treated as
additive. OThink-SRR1 changes refinement architecture and reward design
jointly. PiCA reports aggregate relative improvements; its benchmark-level
outcome-only baselines are unavailable and it is excluded from benchmark-level
effect plots. T and F identify the source table and figure, respectively.
\end{minipage}
\end{table*}

\begin{figure}[!t]
\centering
\includegraphics[width=\columnwidth]{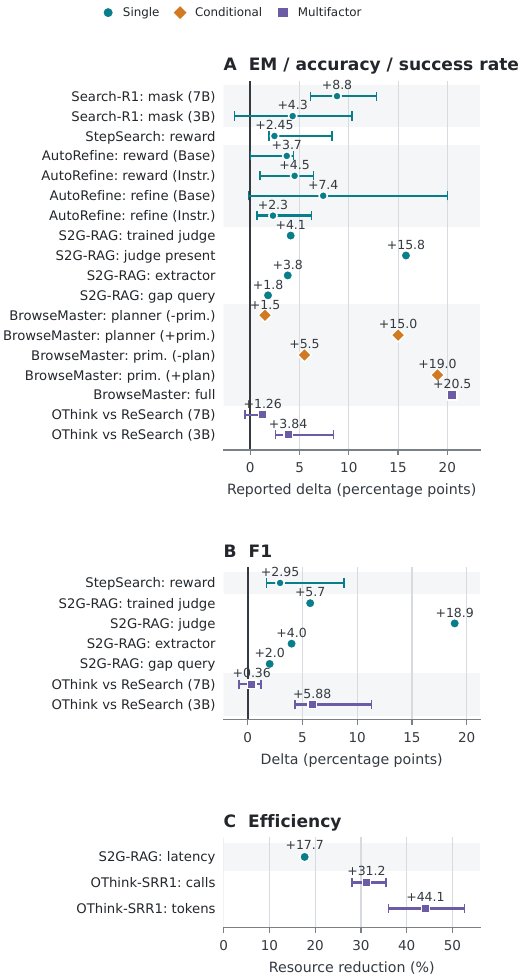}
\caption{Reported within-study deltas from
Table~\ref{tab:controlled-evidence}: (A) EM, accuracy, and success-rate; (B) F1 changes; and (C) resource reductions. Points denote individual results or medians across benchmarks; while whiskers show benchmark ranges, not
confidence intervals. Marker style distinguishes single-component (S),
conditional (C), and multifactor (M) contrasts. Results are not pooled across panels.}
\label{fig:within-study-deltas}
\end{figure}

Figure~\ref{fig:within-study-deltas} makes the direction, magnitude, and
heterogeneity of these reported changes visible without converting them into
standardized effect sizes. Three conclusions emerge. First, acquisition control
and evidence-state management are consequential when isolated: the S2G-RAG
judge and the interacting BrowseMaster planner--primitive modules produce some
of the largest reported gains \cite{pang2025browsemaster,li2026s2grag}. Second,
retrieved-token masking, step-wise rewards, retrieval-aware rewards, and
pivot-based credit assignment generally improve otherwise matched RSA loops
\cite{searchr1,wang2025stepsearch,shi2025autorefine,liu2026pica}. Third,
stateful evidence compression can improve the quality--cost trade-off, as shown
by S2G-RAG's lower latency and OThink-SRR1's reductions in retrieval calls and
input tokens \cite{liang2026othink,li2026s2grag}.

The evidence remains uneven across the taxonomy. The available ablations primarily test retrieval coupling, controller design, evidence-state management, and learning signal; we found no eligible matched comparison that independently isolates knowledge source or grounding strictness. Accordingly, the results provide construct-level evidence that several taxonomy dimensions identify consequential design decisions, but they do not establish that the four axes are sufficient or maximally predictive of RSA-LLM performance.

%% file: IEEE-conference-template-062824/sections/challenges.tex
\section{Challenges and Future Directions}
\label{sec:future}

The proposed taxonomy exposes four complementary research gaps that are obscured when RSA-LLMs are evaluated as a single architectural family.

\noindent\emph{Axis 1: Retrieval coupling.}
Learned search policies remain computationally expensive, and additional retrieval does not necessarily produce useful evidence. Recent work explores simulated retrieval and information-foraging objectives \cite{zerosearch,qian2025scent}, but reliable action-level credit assignment remains unresolved. Retrieval should be optimized as budgeted evidence acquisition, using calibrated stopping criteria and marginal evidence utility relative to search calls, tokens, latency, and monetary cost. Evaluation should consequently report quality--cost Pareto frontiers under matched budgets.

\noindent\emph{Axis 2: Controller design.}
Controller state and coordination logic are frequently embedded in prompts or heuristics, limiting reproducibility and diagnosis. Future systems should expose explicit representations of uncertainty, accumulated evidence, and remaining budget, together with typed interfaces among planner, retriever, integrator, and verifier components. Tree- and graph-based controllers additionally require reproducible pruning, backtracking, and stopping rules. Decision traces and failure logs are particularly important in high-stakes deployments.

\noindent\emph{Axis 3: Knowledge source.}
Comparisons are confounded when systems use different corpora, search APIs, snapshots, or access policies. Evaluations should record these conditions and compare matched controllers across bounded and open-world sources. Unified interfaces are also needed for heterogeneous evidence spanning text, tables, databases, knowledge graphs, and APIs, together with robustness tests for source drift, missing fields, and retrieval-service failure.

\noindent\emph{Axis 4: Grounding strictness.}
Retrieval does not guarantee that generated claims are supported. Future systems should enforce claim-level grounding contracts that require attribution, revision, or abstention when support is insufficient. Evaluation should jointly report citation precision and recall, attribution coverage, unsupported-claim rate, and selective risk \cite{gao-etal-2023-alce}. Because automated verification is itself fallible, adversarial testing and targeted human auditing remain necessary.

%% file: IEEE-conference-template-062824/sections/conclusion.tex
\section{Conclusion}

In this paper, we defined RSA-LLMs as systems that embed state-dependent external evidence acquisition within a closed reasoning--control loop, while treating scheduled retrieval as an adjacent reference point. We introduced four architectural axes: retrieval coupling, controller design, knowledge source, and grounding strictness. We also separated these properties from the learning signals used to optimize system behavior. Beyond descriptive system coding, we synthesized matched within-study ablations and quality--cost contrasts. The available evidence shows that explicit acquisition control, persistent evidence management, and process-aware learning can materially affect answer quality and retrieval efficiency under controlled settings. However, empirical coverage remains uneven for knowledge-source and grounding interventions; the taxonomy should therefore be interpreted as an operational comparison framework rather than a statistically complete predictor of performance. This distinction provides a more reproducible basis for evaluating future RSA systems and identifying where genuinely controlled evidence is still required.

%% file: IEEE-conference-template-062824/backmatter/references.tex
\begingroup
\renewcommand{\footnotesize}{\fontsize{8pt}{8.2pt}\selectfont}
\renewcommand{\IEEEbibitemsep}{-0.60pt}
\bibliographystyle{IEEEtran}
\bibliography{IEEE-conference-template-062824/backmatter/references}
\endgroup

%% file: IEEE-conference-template-062824/backmatter/references.bib
@inproceedings{chen2025researchlearningreasonsearch,
  title     = {ReSearch: Learning to Reason with Search for {LLM}s via Reinforcement Learning},
  author    = {Chen, Mingyang and Sun, Linzhuang and Li, Tianpeng and Sun, Haoze and Zhou, Yijie and Zhu, Chenzheng and Wang, Haofen and Pan, Jeff Z. and Zhang, Wen and Chen, Huajun and Yang, Fan and Zhou, Zenan and Chen, Weipeng},
  booktitle = {NeurIPS},
  volume    = {38},
  year      = {2025}
}

@inproceedings{wang2025stepsearch,
  title     = {StepSearch: Igniting {LLM}s Search Ability via Step-Wise Proximal Policy Optimization},
  author    = {Zheng, Xuhui and An, Kang and Wang, Ziliang and Wang, Yuhang and Wu, Yichao},
  booktitle = {EMNLP},
  pages     = {21805--21830},
  year      = {2025},
  doi       = {10.18653/v1/2025.emnlp-main.1106}
}

@article{rsearch,
  title  = {R-Search: Empowering {LLM} Reasoning with Search via Multi-Reward Reinforcement Learning},
  author = {Zhao, Qingfei and Wang, Ruobing and Xu, Dingling and Zha, Daren and Liu, Limin},
  year   = {2025},
  doi    = {10.48550/arXiv.2506.04185}
}

@article{memsearcher,
  title  = {MemSearcher: Training {LLM}s to Reason, Search and Manage Memory via End-to-End Reinforcement Learning},
  author = {Yuan, Qianhao and others},
  year   = {2025},
  doi    = {10.48550/arXiv.2511.02805}
}

@article{wu2025resumunlockinglonghorizonsearch,
  title  = {ReSum: Unlocking Long-Horizon Search Intelligence via Context Summarization},
  author = {Wu, Xixi and others},
  year   = {2025},
  doi    = {10.48550/arXiv.2509.13313}
}

@inproceedings{qian2025scent,
  title     = {Scent of Knowledge: Optimizing Search-Enhanced Reasoning with Information Foraging},
  author    = {Qian, Hongjin and Liu, Zheng},
  booktitle = {NeurIPS},
  year      = {2025}
}

@article{jiang2025verltoolholisticagenticreinforcement,
  title  = {VerlTool: Towards Holistic Agentic Reinforcement Learning with Tool Use},
  author = {Jiang, Dongfu and others},
  year   = {2025},
  doi    = {10.48550/arXiv.2509.01055}
}

@inproceedings{shi2025autorefine,
  title     = {Search and Refine During Think: Facilitating Knowledge Refinement for Improved Retrieval-Augmented Reasoning},
  author    = {Shi, Yaorui and Li, Sihang and Wu, Chang and Liu, Zhiyuan and Fang, Junfeng and Cai, Hengxing and Zhang, An and Wang, Xiang},
  booktitle = {NeurIPS},
  volume    = {38},
  year      = {2025}
}

@inproceedings{song-etal-2025-rekg-ReKG-MCTS,
  title     = {{R}e{KG}-{MCTS}: Reinforcing {LLM} Reasoning on Knowledge Graphs via Training-Free {M}onte {C}arlo Tree Search},
  author    = {Song, Xiaozhuang and Zhang, Shufei and Yu, Tianshu},
  booktitle = {Findings of ACL},
  year      = {2025},
  pages     = {9288--9306}
}

@inproceedings{JoshiTriviaQA2017,
  title     = {TriviaQA: A Large Scale Distantly Supervised Challenge Dataset for Reading Comprehension},
  author    = {Joshi, Mandar and Choi, Eunsol and Weld, Daniel S. and Zettlemoyer, Luke},
  booktitle = {ACL},
  pages     = {1601--1611},
  year      = {2017}
}

@article{kwiatkowski-etal-2019-natural-questions,
  title   = {Natural Questions: A Benchmark for Question Answering Research},
  author  = {Kwiatkowski, Tom and Palomaki, Jennimaria and Redfield, Olivia and others},
  journal = {TACL},
  volume  = {7},
  pages   = {453--466},
  year    = {2019}
}

@article{trivedi-etal-2022-musique,
  title   = {{M}u{S}i{Q}ue: Multihop Questions via Single-hop Question Composition},
  author  = {Trivedi, Harsh and Balasubramanian, Niranjan and Khot, Tushar and Sabharwal, Ashish},
  journal = {TACL},
  volume  = {10},
  pages   = {539--554},
  year    = {2022}
}

@inproceedings{ho-etal-2020-2WikiMultiHopQA,
  title     = {Constructing a Multi-hop {QA} Dataset for Comprehensive Evaluation of Reasoning Steps},
  author    = {Ho, Xanh and Nguyen, Anh-Khoa Duong and Sugawara, Saku and Aizawa, Akiko},
  booktitle = {COLING},
  year      = {2020},
  pages     = {6609--6625}
}

@inproceedings{patil2025bfcl,
  title     = {The Berkeley Function Calling Leaderboard ({BFCL}): From Tool Use to Agentic Evaluation of Large Language Models},
  author    = {Patil, Shishir G. and Mao, Huanzhi and Yan, Fanjia and Ji, Charlie Cheng-Jie and Suresh, Vishnu and Stoica, Ion and Gonzalez, Joseph E.},
  booktitle = {ICML},
  year      = {2025}
}

@inproceedings{mialon2024gaia,
title={{GAIA}: a benchmark for General {AI} Assistants},
author={Gr{\'e}goire Mialon and Cl{\'e}mentine Fourrier and Thomas Wolf and Yann LeCun and Thomas Scialom},
booktitle = {ICLR},
year={2024},
}

@article{wei2025browsecompsimplechallengingbenchmark,
  title  = {BrowseComp: A Simple Yet Challenging Benchmark for Browsing Agents},
  author = {Wei, Jason and Sun, Zhiqing and Papay, Spencer and others},
  journal = {arXiv:2504.12516},
  year   = {2025},
  doi    = {10.48550/arXiv.2504.12516}
}

@inproceedings{yang-etal-2018-hotpotqa,
  title     = {HotpotQA: A Dataset for Diverse, Explainable Multi-hop Question Answering},
  author    = {Yang, Zhilin and others},
  booktitle = {EMNLP},
  pages     = {2369--2380},
  year      = {2018}
}

@article{besta2025demystifying,
  title   = {Demystifying Chains, Trees, and Graphs of Thoughts},
  author  = {Besta, Maciej and others},
  journal = {IEEE TPAMI},
  year    = {2025}
}

@inproceedings{barnett2024seven,
  title     = {Seven Failure Points When Engineering a Retrieval-Augmented Generation System},
  author    = {Barnett, Scott and others},
  booktitle = {IEEE/ACM CAIN},
  year      = {2024},
  pages     = {194--199}
}

@inproceedings{wei2022chain,
  title     = {Chain-of-Thought Prompting Elicits Reasoning in Large Language Models},
  author    = {Wei, Jason and others},
  booktitle = {NeurIPS},
  year      = {2022}
}

@inproceedings{kojima2022zeroshot,
  title     = {Large Language Models Are Zero-Shot Reasoners},
  author    = {Kojima, Takeshi and others},
  booktitle = {NeurIPS},
  year      = {2022}
}

@inproceedings{petroni2019lmkb,
    title = "Language Models as Knowledge Bases?",
    author = {Petroni, Fabio  and
      Rockt{\"a}schel, Tim  and
      Riedel, Sebastian  and
      Lewis, Patrick  and
      Bakhtin, Anton  and
      Wu, Yuxiang  and
      Miller, Alexander},
    year = "2019",
    booktitle = “EMNLP-IJCNLP”,
    url = "https://aclanthology.org/D19-1250/",
    doi = "10.18653/v1/D19-1250",
    pages = "2463--2473"
}

@inproceedings{lewis2020rag,
  title     = {Retrieval-Augmented Generation for Knowledge-Intensive NLP Tasks},
  author    = {Lewis, Patrick and Perez, Ethan and Piktus, Aleksandra and Petroni, Fabio and Karpukhin, Vladimir and Goyal, Naman and K{\"u}ttler, Heinrich and Lewis, Mike and Yih, Wen-tau and Rockt{\"a}schel, Tim and Riedel, Sebastian and Kiela, Douwe},
  booktitle = {NeurIPS},
  year      = {2020}
}

@inproceedings{lin2022truthfulqa,
  title     = {TruthfulQA: Measuring How Models Mimic Human Falsehoods},
  author    = {Lin, Stephanie and Hilton, Jacob and Evans, Owain},
  booktitle = {ACL},
  pages     = {3214--3252},
  year      = {2022}
}

@inproceedings{wang2023selfconsistency,
  title     = {Self-Consistency Improves Chain-of-Thought Reasoning in Language Models},
  author    = {Wang, Xuezhi and Wei, Jason and Schuurmans, Dale and Le, Quoc V. and Chi, Ed H. and Narang, Sharan and Chowdhery, Aakanksha and Zhou, Denny},
  booktitle = {ICLR},
  year      = {2023}
}

@inproceedings{press2023measuring,
  title     = {Measuring and Narrowing the Compositionality Gap in Language Models},
  author    = {Press, Ofir and Zhang, Muru and Min, Sewon and Schmidt, Ludwig and Smith, Noah A. and Lewis, Mike},
  booktitle = {Findings of EMNLP},
  pages     = {5687--5711},
  year      = {2023}
}

@misc{fan2024surveyragmeetingllms,
  title  = {A Survey on RAG Meeting LLMs: Towards Retrieval-Augmented Large Language Models},
  author = {Fan, Wenqi and Ding, Yujuan and Ning, Liangbo and Wang, Shijie and Li, Hengyun and Yin, Dawei and Chua, Tat-Seng and Li, Qing},
  year   = {2024},
  doi    = {10.48550/arXiv.2405.06211}
}

@inproceedings{li-etal-2025-survey,
  title     = {A Survey of {RAG}-Reasoning Systems in Large Language Models},
  author    = {Li, Yangning and Zhang, Weizhi and Yang, Yuyao and others},
  booktitle = {Findings of EMNLP},
  year      = {2025},
  doi       = {10.18653/v1/2025.findings-emnlp.648}
}

@inproceedings{yao2023react,
  title     = {ReAct: Synergizing Reasoning and Acting in Language Models},
  author    = {Yao, Shunyu and Zhao, Jeffrey and Yu, Dian and Du, Nan and Shafran, Izhak and Narasimhan, Karthik and Cao, Yuan},
  booktitle = {ICLR},
  year      = {2023}
}

@misc{li2025searcho1,
  title  = {Search-o1: Agentic Search-Enhanced Large Reasoning Models},
  author = {Li, Xiaoxi and Dong, Guanting and Jin, Jiajie and Zhang, Yuyao and Zhou, Yujia and Zhu, Yutao and Zhang, Peitian and Dou, Zhicheng},
  year   = {2025},
  doi    = {10.48550/arXiv.2501.05366}
}

@article{song2025r1searcher,
  title  = {R1-Searcher: Incentivizing the Search Capability in {LLM}s via Reinforcement Learning},
  author = {Song, Huatong and others},
  year   = {2025},
  doi    = {10.48550/arXiv.2503.05592}
}

@article{yao2023tree,
  title  = {Tree of Thoughts: Deliberate Problem Solving with Large Language Models},
  author = {Yao, Shunyu and Yu, Dian and Zhao, Jeffrey and Shafran, Izhak and Griffiths, Thomas L. and Cao, Yuan and Narasimhan, Karthik},
  year   = {2023},
  doi    = {10.48550/arXiv.2305.10601}
}

@article{browne2012mcts,
  title   = {A Survey of Monte Carlo Tree Search Methods},
  author  = {Browne, Cameron B. and others},
  journal = {IEEE Trans. Comput. Intell. AI Games},
  volume  = {4},
  number  = {1},
  pages   = {1--43},
  year    = {2012},
  doi = {10.1109/TCIAIG.2012.2186810}
}

@inproceedings{sun2024tog,
title={Think-on-Graph: Deep and Responsible Reasoning of Large Language Model on Knowledge Graph},
author={Jiashuo Sun and Chengjin Xu and Lumingyuan Tang and Saizhuo Wang and Chen Lin and Yeyun Gong and Lionel Ni and Heung-Yeung Shum and Jian Guo},
booktitle={ICLR},
year={2024},
}

@inproceedings{jin-etal-2024-graph,
  title     = {Graph Chain-of-Thought: Augmenting Large Language Models by Reasoning on Graphs},
  author    = {Jin, Bowen and Xie, Chulin and Zhang, Jiawei and Roy, Kashob Kumar and Zhang, Yu and Li, Zheng and Li, Ruirui and Tang, Xianfeng and Wang, Suhang and Meng, Yu and Han, Jiawei},
  booktitle = {Findings of ACL},
  year      = {2024},
  pages     = {163--184}
}

@inproceedings{li-etal-2024-graphreader,
  title     = {GraphReader: Building Graph-based Agent to Enhance Long-Context Abilities of Large Language Models},
  author    = {Li, Shilong and He, Yancheng and Guo, Hangyu and others},
  journal   = {arXiv preprint arXiv:2406.14550},
  year      = {2024},
  doi       = {10.48550/arXiv.2406.14550}
}

@inproceedings{izacard2021fid,
  title     = {Leveraging Passage Retrieval with Generative Models for Open-Domain Question Answering},
  author    = {Izacard, Gautier and Grave, Edouard},
  booktitle = {EACL},
  year      = {2021}
}

@inproceedings{asai2024selfrag,
author={Asai, Akari and Wu, Zeqiu and Wang, Yizhong and Sil, Avirup and Hajishirzi, Hannaneh},
title={Self-{RAG}: Learning to Retrieve, Generate, and Critique through Self-Reflection},
url={https://openreview.net/forum?id=hSyW5go0v8},
booktitle={ICLR},
year={2024}
}

@inproceedings{gao-etal-2023-alce,
  title     = {Enabling Large Language Models to Generate Text with Citations},
  author    = {Gao, Tianyu and Yen, Howard and Yu, Jiatong and Chen, Danqi},
  booktitle = {EMNLP},
  year      = {2023}
}

@article{menick2022gophercite,
  title   = {Teaching language models to support answers with verified quotes},
  author  = {Menick, Jacob and others},
  journal = {arXiv preprint arXiv:2203.11147},
  year    = {2022}
}

@inproceedings{thorne-etal-2018-fever,
  title     = {{FEVER}: a Large-scale Dataset for Fact Extraction and {VER}ification},
  author    = {Thorne, James and others},
  booktitle = {NAACL},
  pages     = {809--819},
  year      = {2018}
}

@article{karpas2022mrkl,
  title   = {{MRKL} Systems: A Modular, Neuro-Symbolic Architecture that Combines Large Language Models, External Knowledge Sources and Discrete Reasoning},
  author  = {Karpas, Ehud and others},
  journal = {arXiv:2205.00445},
  year    = {2022},
  doi = {10.48550/arXiv.2205.00445}
}

@inproceedings{gao-etal-2023-rarr,
  title     = {RARR: Researching and Revising What Language Models Say, Using Language Models},
  author    = {Gao, Luyu and Dai, Zhuyun and Pasupat, Panupong and Chen, Anthony and Chaganty, Arun Tejasvi and Fan, Yicheng and Zhao, Vincent Y. and Lao, Ni and Lee, Hongrae and Juan, Da-Cheng and Guu, Kelvin},
  booktitle = {ACL},
  year      = {2023}
}

@inproceedings{searchr1,
  title     = {{Search-R1}: Training {LLM}s to Reason and Leverage Search Engines with Reinforcement Learning},
  author    = {Jin, Bowen and Zeng, Hansi and Yue, Zhenrui and Yoon, Jinsung and Arik, Sercan O. and Wang, Dong and Zamani, Hamed and Han, Jiawei},
  booktitle = {COLM},
  year      = {2025}
}

@article{zerosearch,
  title   = {{ZeroSearch}: Incentivize the Search Capability of {LLM}s without Searching},
  author  = {Sun, Hao and others},
  journal = {arXiv preprint arXiv:2505.04588},
  year    = {2025},
  doi     = {10.48550/arXiv.2505.04588}
}

@article{nakano2021webgpt,
  title   = {{WebGPT}: Browser-assisted Question-Answering with Human Feedback},
  author  = {Nakano, Reiichiro and Hilton, Jacob and Balaji, Suchir and others},
  journal = {arXiv preprint arXiv:2112.09332},
  year    = {2021},
  doi     = {10.48550/arXiv.2112.09332}
}

@inproceedings{shen2023hugginggpt,
  title     = {{HuggingGPT}: Solving {AI} Tasks with {ChatGPT} and its Friends in Hugging Face},
  author    = {Shen, Yongliang and Song, Kaitao and Tan, Xu and Li, Dongsheng and Lu, Weiming and Zhuang, Yueting},
  booktitle = {NeurIPS},
  year      = {2023}
}

@article{du2023multiagentdebate,
  title        = {Improving Factuality and Reasoning in Language Models through Multiagent Debate},
  author       = {Du, Yilun and Li, Shuang and Torralba, Antonio and Tenenbaum, Joshua B. and Mordatch, Igor},
  journal      = {arXiv preprint arXiv:2305.14325},
  year         = {2023}
}

@inproceedings{liang2024mad,
    title = "Encouraging Divergent Thinking in Large Language Models through Multi-Agent Debate",
    author = "Liang, Tian  and
      He, Zhiwei  and
      Jiao, Wenxiang  and
      Wang, Xing  and
      Wang, Yan  and
      Wang, Rui  and
      Yang, Yujiu  and
      Shi, Shuming  and
      Tu, Zhaopeng",
    year = "2024",
    url = "https://aclanthology.org/2024.emnlp-main.992/",
    booktitle = "EMNLP",
    doi = "10.18653/v1/2024.emnlp-main.992",
    pages = "17889--17904",
}

@article{besta2023graphofthoughts,
  title        = {Graph of Thoughts: Solving Elaborate Problems with Large Language Models},
  author       = {Besta, Maciej and others},
  booktitle = {AAAI},
  volume={38},
  number={16},
  pages={17682--17690},
  year={2024},
  doi       = {10.1609/aaai.v38i16.29720}
}

@article{liu2024lostmiddle,
  author = "Liu, Nelson F.  and
      Lin, Kevin  and
      Hewitt, John  and
      Paranjape, Ashwin  and
      Bevilacqua, Michele  and
      Petroni, Fabio  and
      Liang, Percy",
  title   = {Lost in the Middle: How Language Models Use Long Contexts},
  journal = "TACL",
  volume  = {12},
  pages   = {157--173},
  year    = {2024}
}

@inproceedings{li-etal-2023-apibank,
  title     = {{API}-Bank: A Comprehensive Benchmark for Tool-Augmented {LLM}s},
  author    = {Li, Minghao and Zhao, Yingxiu and Yu, Bowen and Song, Feifan and Li, Hangyu and Yu, Haiyang and Li, Zhoujun and Huang, Fei and Li, Yongbin},
  booktitle = {EMNLP},
  year      = {2023},
  pages     = {3102--3116}
}

@misc{schick2023toolformer,
  title  = {Toolformer: Language Models Can Teach Themselves to Use Tools},
  author = {Schick, Timo and Dwivedi-Yu, Jane and Dess{\`\i}, Roberto and others},
  journal = {NeurIPS},
  year   = {2023},
  doi    = {10.48550/arXiv.2302.04761}
}

@inproceedings{zhu2021adaptive,
  title={Adaptive information seeking for open-domain question answering},
  author={Zhu, Yunchang and Pang, Liang and Lan, Yanyan and Shen, Huawei and Cheng, Xueqi},
  booktitle={EMNLP},
  year={2021}
}

@inproceedings{jiang2024resp,
author = {Jiang, Zhouyu and Sun, Mengshu and Liang, Lei and Zhang, Zhiqiang},
title = {Retrieve, Summarize, Plan: Advancing Multi-hop Question Answering with an Iterative Approach},
year = {2025},
publisher = {ACM},
url = {https://doi.org/10.1145/3701716.3716889},
doi = {10.1145/3701716.3716889},
pages = {1677–1686},
booktitle = {Companion Proc. WWW}
}

@article{nahid2025prism,
  title   = {PRISM: Agentic Retrieval with LLMs for Multi-Hop Question Answering},
  author  = {Nahid, Md Mahadi Hasan and Rafiei, Davood},
  journal = {arXiv preprint arXiv:2510.14278},
  year    = {2025}
}

@article{li2025knowtrace,
  title   = {KnowTrace: Bootstrapping Iterative Retrieval-Augmented Generation with Structured Knowledge Tracing},
  author  = {Li, Rui and Dai, Quanyu and Zhang, Zeyu and Chen, Xu and Dong, Zhenhua and Wen, Ji-Rong},
  journal = {arXiv preprint arXiv:2505.20245},
  year    = {2025}
}

@article{singh2025agenticrag,
  title   = {Agentic Retrieval-Augmented Generation: A Survey on Agentic {RAG}},
  author={Aditi Singh and Abul Ehtesham and Saket Kumar and Tala Talaei Khoei and Athanasios V. Vasilakos},
  journal = {arXiv preprint arXiv:2501.09136},
  year    = {2025}
}

@article{qin2024toollearning,
  title   = {Tool Learning with Foundation Models},
  author  = {Qin, Yujia and Hu, Shengding and Lin, Yankai and others},
  journal = {ACM Computing Surveys},
  volume  = {57},
  number  = {4},
  pages   = {1--40},
  year    = {2024},
  doi     = {10.1145/3704435}
}

@inproceedings{krishna2025frames,
  title     = {Fact, Fetch, and Reason: A Unified Evaluation of Retrieval-Augmented Generation},
  author    = {Krishna, Satyapriya and Krishna, Kalpesh and Mohananey, Anhad and Schwarcz, Steven and Stambler, Adam and Upadhyay, Shyam and Faruqui, Manaal},
  booktitle = {NAACL-HLT},
  pages     = {4745--4759},
  year      = {2025},
  doi       = {10.18653/v1/2025.naacl-long.243}
}

@inproceedings{pham2026sealqa,
  title     = {{SEALQA}: Raising the Bar for Reasoning in Search-Augmented Language Models},
  author    = {Pham, Thinh and Nguyen, Nguyen and Zunjare, Pratibha and Chen, Weiyuan and Tseng, Yu-Min and Vu, Tu},
  booktitle = {ICLR},
  year      = {2026}
}

@article{li2026sgrbench,
  title   = {{SGR}-Bench: Benchmarking Search Agents on State-Gated Retrieval},
  author  = {Li, Ningyuan and Shen, Haiyang and Liu, Mugeng and Han, Yudong and Shi, Zhuofan and Xie, Sixiong and Ma, Yun},
  journal = {arXiv preprint arXiv:2605.22219},
  year    = {2026},
  doi     = {10.48550/arXiv.2605.22219}
}

@article{wang2026merrin,
  title   = {{MERRIN}: A Benchmark for Multimodal Evidence Retrieval and Reasoning in Noisy Web Environments},
  author  = {Wang, Han and Wan, David and Lee, Hyunji and Pham, Thinh and Cankosyan, Mikaela and Chen, Weiyuan and Stengel-Eskin, Elias and Vu, Tu and Bansal, Mohit},
  journal = {arXiv preprint arXiv:2604.13418},
  year    = {2026},
  doi     = {10.48550/arXiv.2604.13418}
}

@article{zhou2025browsecompzh,
  title   = {{BrowseComp-ZH}: Benchmarking Web Browsing Ability of Large Language Models in Chinese},
  author  = {Zhou, Peilin and Leon, Bruce and Ying, Xiang and Zhang, Can and Shao, Yifan and Ye, Qichen and Chong, Dading and Jin, Zhiling and Xie, Chenxuan and Cao, Meng and Gu, Yuxin and Hong, Sixin and Ren, Jing and Chen, Jian and Liu, Chao and Hua, Yining},
  journal = {arXiv preprint arXiv:2504.19314},
  year    = {2025},
  doi     = {10.48550/arXiv.2504.19314}
}

@inproceedings{wu2025webwalker,
  title     = {WebWalker: Benchmarking {LLM}s in Web Traversal},
  author    = {Wu, Jialong and Yin, Wenbiao and Jiang, Yong and Wang, Zhenglin and Xi, Zekun and Fang, Runnan and Zhang, Linhai and He, Yulan and Zhou, Deyu and Xie, Pengjun and Huang, Fei},
  booktitle = {ACL},
  pages     = {10290--10305},
  year      = {2025},
  doi       = {10.18653/v1/2025.acl-long.508}
}

@article{li2026deepresearchbench2,
  title   = {DeepResearch Bench II: Diagnosing Deep Research Agents via Rubrics from Expert Report},
  author  = {Li, Ruizhe and Du, Mingxuan and Xu, Benfeng and Zhu, Chiwei and Wang, Xiaorui and Mao, Zhendong},
  journal = {arXiv preprint arXiv:2601.08536},
  year    = {2026},
  doi     = {10.48550/arXiv.2601.08536}
}

@article{liang2026othink,
  title   = {{OThink-SRR1}: Search, Refine and Reasoning with Reinforced Learning for Large Language Models},
  author  = {Liang, Haijian and Niu, Zenghao and Wu, Junjie and Zhang, Changwang and Zhou, Wangchunshu and Wang, Jun},
  journal = {arXiv preprint arXiv:2604.19766},
  year    = {2026},
  doi     = {10.48550/arXiv.2604.19766}
}

@article{pang2025browsemaster,
  title   = {{BrowseMaster}: Towards Scalable Web Browsing via Tool-Augmented Programmatic Agent Pair},
  author  = {Pang, Xianghe and Tang, Shuo and Ye, Rui and Du, Yuwen and Du, Yaxin and Chen, Siheng},
  journal = {arXiv preprint arXiv:2508.09129},
  year    = {2025},
  doi     = {10.48550/arXiv.2508.09129}
}

@inproceedings{oswald2026modelspace,
  title     = {Model Space Reasoning as Search in Feedback Space for Planning Domain Generation},
  author    = {Oswald, James and Obolensky, Daniel and Varha, Volodymyr and Dragovic, Vasilije and Srinivas, Kavitha and Kokel, Harsha and Katz, Michael and Sohrabi, Shirin},
  booktitle = {ICLR Workshop on World Models},
  year      = {2026},
  doi       = {10.48550/arXiv.2604.08712}
}

@inproceedings{trivedi2023ircot,
  title     = {Interleaving Retrieval with Chain-of-Thought Reasoning for Knowledge-Intensive Multi-Step Questions},
  author    = {Trivedi, Harsh and Balasubramanian, Niranjan and Khot, Tushar and Sabharwal, Ashish},
  booktitle = {ACL},
  pages     = {10014--10037},
  year      = {2023},
  doi       = {10.18653/v1/2023.acl-long.557}
}

@inproceedings{li2026s2grag,
  author    = {Li, Minghan and Zou, Junjie and Lv, Xinxuan and Zhang, Chao and Zhou, Guodong},
  title     = {{S2G-RAG}: Structured Sufficiency and Gap Judging for Iterative Retrieval-Augmented {QA}},
  booktitle = {Proceedings of the 64th Annual Meeting of the Association for Computational Linguistics (Volume 1: Long Papers)},
  pages     = {25846--25862},
  year      = {2026},
  doi       = {10.18653/v1/2026.acl-long.1185},
  url       = {https://aclanthology.org/2026.acl-long.1185/}
}

@article{liu2026pica,
  title   = {{PiCA}: Pivot-Based Credit Assignment for Search Agentic Reinforcement Learning},
  author  = {Liu, Dongyi and Niu, Yifan and Wang, Qinwen and Xiao, Han and Li, Jia},
  journal = {arXiv preprint arXiv:2605.09287},
  year    = {2026},
  doi     = {10.48550/arXiv.2605.09287}
}
